\documentclass[a4paper,fleqn,usenatbib]{mnras}

\usepackage{newtxtext,newtxmath}
\usepackage[T1]{fontenc}
\usepackage{ae,aecompl}

\usepackage{graphicx}	
\usepackage{amsmath}	
\usepackage{amssymb}	
\usepackage{adjustbox}
\usepackage{xspace}
\usepackage{natbib}
\setcitestyle{notesep={}}
\usepackage{etoolbox}
\makeatletter
\patchcmd\@combinedblfloats{\box\@outputbox}{\unvbox\@outputbox}{}{%
  \errmessage{\noexpand\@combinedblfloats could not be patched}%
}%
\makeatother
\usepackage{cleveref}
\crefname{figure}{Figure}{Figures}
\crefname{section}{Section}{Sections}
\crefname{table}{Table}{Tables}
\crefname{appendix}{Appendix}{Appendices}

\newcommand{\Msol}{M$_{\odot}$\xspace}

\newcommand{\squotes}[1]{\lq {#1}\rq\xspace}
\newcommand{\dquotes}[1]{\lq\lq {#1}\rq\rq\xspace}
\newcommand{\eagle}{{\sc eagle}\xspace}

\newcommand{\subfind}{{\sc subfind}\xspace}

\newcommand{\Msolyr}{M$_{\odot}$~yr$^{-1}$\xspace}

\newcommand{\M}[1]{$M_{\mathrm{#1}}$\xspace}

\newcommand{\smm}{submm\xspace}
\newcommand{\fsmm}{$S_{850 \mu \mathrm{m}}$\xspace}
\newcommand{\smmfaint}{Submm-Faint\xspace}

\newcommand{\smmbright}{\fsmm $\geq 1$~mJy\xspace}
\newcommand{\mergerratio}{$M_{*,1} / M_{*,2}$\xspace}

\title[Sub-millimeter galaxies in \eagle]{The nature of sub-millimeter and highly star-forming galaxies in the \eagle simulation}

\author[S. McAlpine et al.]{Stuart McAlpine,$^{1,2}$\thanks{E-mail:
stuart.mcalpine@helsinki.fi} Ian Smail,$^{3}$ Richard G. Bower,$^{1}$ Mark A. Swinbank,$^{3}$
\newauthor
James W. Trayford,$^{4}$ Tom Theuns,$^{1}$ Maarten Baes,$^{5}$ Peter Camps,$^{5}$ Robert A. Crain,$^{6}$
\newauthor
and Joop Schaye$^{4}$
\\ $^{1}$Institute for Computational Cosmology, Department of Physics, Durham University, South Road, Durham, DH1 3LE, UK\\ $^{2}$Department of Physics, University of Helsinki, Gustaf H\"{a}llstr\"{o}min katu 2a P.O. Box 64, FI-00014 University of Helsinki, Finland\\ $^{3}$Centre for Extragalactic
Astronomy, Department of Physics, Durham University, South Road, Durham DH1
3LE, UK\\ $^{4}$Leiden Observatory,
Leiden University, P.O. Box 9513, 2300 RA Leiden, the Netherlands\\ $^{5}$Sterrenkundig Observatorium, Universiteit Gent, Krijgslaan 281, B-9000 Gent, Belgium\\ $^{6}$Astrophysics Research Institute, Liverpool John Moores University, 146 Brownlow Hill, Liverpool L3 5RF, UK}

\date{Accepted XXX. Received YYY; in original form ZZZ}

\pubyear{2017}

\begin{document}
\label{firstpage}
\pagerange{\pageref{firstpage}--\pageref{lastpage}}
\maketitle

\begin{abstract}
We exploit \eagle, a cosmological hydrodynamical simulation, to reproduce the selection of the observed sub-millimeter (\smm) galaxy population by selecting the model galaxies at $z \geq 1$ with mock \smm fluxes \fsmm $\geq 1$~mJy (computed in post processing using radiative transfer techniques). There is a reasonable agreement between the galaxies within this sample and the properties of the observed \smm population, such as their star formation rates (SFRs) at $z<3$, redshift distribution and many integrated galaxy properties. We find that the bulk of the \smmbright model population is at $z \approx 2.5$, and that they are massive galaxies ($M_* \sim 10^{11}$~\Msol) with high dust masses ($M_{\mathrm{dust}} \sim 10^{8}$~\Msol), gas fractions ($f_{\mathrm{gas}} \approx 50$\%) and SFRs ($\dot M_* \approx 100$~\Msolyr). They have major and minor merger fractions similar to the general population, suggesting that mergers are not the primary driver of the model \smm galaxies. Instead, the \smmbright model galaxies yield high SFRs primarily because they maintain a 
significant gas reservoir as a result of hosting an undermassive black hole. In addition, we find that not all \squotes{highly star-forming} ($\dot M_* \geq 80$~\Msolyr) \eagle galaxies have \smm fluxes \smmbright. Thus, we investigate the nature of $z \geq 1$ highly star-forming \squotes{\smmfaint} galaxies (i.e., $\dot M_* \geq 80$~\Msolyr but \fsmm $< 1$~mJy). We find they are similar to the model \smm galaxies; being gas rich and hosting undermassive black holes, however they are typically lower mass ($M_* \sim 10^{10}$~\Msol) and are at higher redshifts ($z>4$). These typically higher-redshift galaxies show stronger evidence for having been triggered by major mergers, and critically, they are likely missed by current submm surveys due to their higher dust temperatures. This suggests a potentially even larger contribution to the SFR density at $z \gtrsim 3$ from dust-obscured systems than implied by current observations.

\end{abstract}


\section{Introduction}
\label{sect:introdution}

Sub-millimeter (\smm) galaxies (SMGs) are a population of high-redshift galaxies ($z \approx 1$--5) which are inferred to have high star-formation rates (SFRs, $\dot M_* \sim$ 100~\Msolyr) and significant dust masses ($M_{\mathrm{dust}} \sim 10^{8}$~\Msol). The luminous \smm emission arises from the reprocessing of the ultraviolet (UV) light from young stars by obscuring dust \citep[see][ for a review]{Casey2014}. SMGs are relatively rare, with a number density of $\sim 10^{-5}$ cMpc$^{-3}$ at $z\approx$\,2--3 \citep[e.g.,][]{Chapman2005,Simpson2014}. However, this population is particularly interesting due to the apparently high levels of star formation and the many open questions that remain to be answered about their formation and evolution. For example:  What triggers such extreme SFRs, mergers or secular disk instabilities?  How long can these galaxies form stars at such apparently high rates, and what terminates it?  How do they evolve following the starburst episode? What are their descendants at $z \approx 0$? Many of these questions cannot be answered directly by the observations, however, by combining the observational data with advanced numerical simulations, they can begin to be addressed.

Previous comparisons of models with observations of this galaxy population have highlighted an apparent problem with current theories of galaxy formation \citep[e.g.,][]{Baugh2005,Swinbank2008,Dave2010,Hayward2013}.  These models generally have difficulty in matching the observed SFRs or \smm number counts of SMGs while maintaining agreement with other observations of the wider galaxy population. A number of solutions have been proposed for this discrepancy; including a top-heavy initial mass function (IMF), the unaccounted for blending of multiple galaxies into a single \smm source in the observations, or the inability of hydrodynamical simulations to adequately resolve intense starbursts (leading to an underestimate in the SFR). However, recently the necessity of a radical solution to the discrepancy has perhaps weakened since higher resolution observational data has become available \citep[e.g.,][]{Karim2013,Simpson2015,Cowie2018}, which now resolves the blending of multiple SMGs into a single source \citep[e.g.,][]{Cowley2015}.


In this study we investigate the prevalence of \smm and highly star-forming galaxies in the largest simulation of the \eagle project \citep{Crain2015,Schaye2015,McAlpine2015}. This simulation was calibrated to reproduce the observed galaxy stellar mass function, the sizes of galaxies and the black hole--stellar mass relation at $z \approx 0.1$. Many other observed galaxy trends have also shown broad agreement with observations, both locally \citep[e.g.,][]{Schaye2015,Trayford2015,Lagos2015,Bahe2016} and at higher redshift \citep[e.g.,][]{Furlong2015,Rahmati2015,Furlong2017,McAlpine2017}. Predictions from the simulation for the full model galaxy population are reasonably representative of the observed Universe. As a result, this simulation presents an interesting testbed for more extreme populations, such as those galaxies with the highest SFRs and the SMG population. Observationally it is claimed that these two populations strongly overlap. We can test this in our model, by selecting the samples independently.

This paper is organised as follows: In  \S\ref{sect:eagle} we provide a brief overview of the \eagle simulation and discuss the sample selections. In \S\ref{sect:obs_comparison} we compare our sample of \smm-selected model galaxies to the observed \smm population. We then examine the nature and triggering of the highly star-forming/\smm model galaxies in \S\ref{sect:nature_of_smm} and their descendants at $z=0$ in \S\ref{sect:descendants}. Finally, in \S\ref{sect:conclusions} we present a summary and our concluding remarks.

\section{The \eagle simulation}
\label{sect:eagle}

\eagle \citep[\dquotes{Evolution and Assembly of GaLaxies and their Environment},][]{Schaye2015,Crain2015} \footnote{\url{www.eaglesim.org}}\textsuperscript{,}\footnote{The galaxy and halo catalogues of the simulation suite, as well as the particle data, are publicly available at \url{http://www.eaglesim.org/database.php} \citep{McAlpine2015,EAGLE2017}.} is a suite of cosmological smoothed particle hydrodynamics (SPH) simulations that cover a range of periodic volumes, numerical resolutions and physical models. To incorporate the processes that operate below the simulation resolution a series of \squotes{subgrid} prescriptions are implemented, namely: radiative cooling and photo-ionisation heating \citep{Wiersma2009a}; star formation \citep{Schaye2008}, stellar mass loss \citep{Wiersma2009b} and stellar feedback \citep{DallaVecchia_Schaye2012}; black hole growth via accretion and mergers, and black hole feedback \citep{Springel2005a,Schaye2015,RosasGuevara2016}. The free parameters of these models are calibrated to reproduce the observed galaxy stellar mass function, galaxy sizes and black hole mass--bulge mass relation at $z \approx 0.1$.  A full description of the simulation and the calibration strategy can be found in \citet{Schaye2015} and \citet{Crain2015}, respectively. 

For this study we are interested in the most strongly star-forming ($\dot M_{*} \geq 80$ \Msolyr) and \smm luminous (i.e., those with mock fluxes \fsmm $\geq 1$~mJy) galaxies, and therefore restrict our study to the largest simulation, Ref-L0100N1504, which contains the greatest number of these objects.  This simulation is a cubic periodic volume 100 comoving megaparsecs (cMpc) on each side, sampled by $1504^{3}$ dark matter particles of mass $9.7 \times 10^{6}$~\Msol and an equal number of baryonic particles with an initial mass of $1.8 \times 10^{6}$~\Msol.  The subgrid parameters are those of the \eagle reference model, described fully by \citet{Schaye2015}. The simulation adopts a flat $\Lambda$CDM cosmogony with parameters inferred from analysis of {\it Planck} data \citep{Planck2013}: $\Omega_\Lambda = $\,0.693, $\Omega_{\rm m} = $\,0.307, $\Omega_{\rm b} =$\,0.048, $\sigma_8 =$\,0.8288, $n_{\rm s} = $\,0.9611 and $H_0 = $\,67.77\,km\,s$^{-1}$\,Mpc$^{-1}$. A \citet{Chabrier2003} stellar initial mass function (IMF) is adopted. Unless otherwise stated, error estimates are from a bootstrap analysis.  

\subsection{Simulation output}
\label{sect:sim_output}

The complete state of the simulation is stored at 29 intervals between redshifts $z=20$ and $z=0$, which we refer to as \squotes{snapshots}. In order to produce a halo and galaxy catalogue from these, the dark matter structure finding algorithm \dquotes{Friends of Friends} and the substructure finding algorithm \subfind \citep{Springel2001,Dolag2009} are performed.

Halo mass, \M{200}, is defined as the total mass enclosed within $r_{\mathrm{200}}$, the radius at which the mean enclosed density is 200 times the critical density of the Universe. Galaxy mass, \M{*}, is defined as the total stellar content bound to a subhalo within a spherical aperture with radius 30~proper kiloparsecs (pkpc), as per \citet{Schaye2015}. The sizes of galaxies are quoted in physical units, unless stated otherwise.

\subsubsection{Merger trees}
\label{sect:merger_trees}

Galaxies are tracked from their creation to the present day using a merger tree. Due to the hierarchical build-up, each galaxy today has many progenitors, therefore the history of each galaxy is considered from the reference frame of their \squotes{main progenitor}, the branch of the galaxy's full merger tree that contains the greatest total mass \citep[see][ for full details]{Qu2017}.

Using these merger trees, the merger history of a galaxy can be established. We define the completion time of a galaxy merger with the main-progenitor galaxy to be the cosmic time at the first simulation output where the two galaxies are now identified as a single bound object by the \subfind algorithm. Mergers are classified by their stellar mass ratio, \mergerratio, where $M_{*,2}$ is the mass of the most massive member of the binary. They are considered major if \mergerratio $\geq \frac{1}{4}$, minor if $\frac{1}{10} \leq$ \mergerratio $< \frac{1}{4}$ and either major or minor if \mergerratio $\geq \frac{1}{10}$. To account for the stellar stripping that occurs during the later stages of the interaction, the stellar mass ratio is computed when the in-falling galaxy had its maximum mass \citep[e.g.,][]{Rodriguez-Gomez2015,Qu2017}. Additionally, mergers are only considered \squotes{resolved} when $M_{*,2} \geq 10^{8}$ \Msol ($\approx 100$ stellar particles), but this doesn't effect the results here, where we focus on much more massive galaxies.

\subsubsection{Obtaining accurate star-formation histories of \eagle galaxies}
\label{sect:sfr_from_particles}

The reported \squotes{instantaneous} SFR of an \eagle galaxy is computed from the current state of the galaxy's associated gas particles. However, it is also possible to reverse engineer the SFR of a galaxy at a given time from the galaxy's associated stellar particles at $z=0$\footnote{Note that only the stellar particles born within the main-progenitor (see \cref{sect:merger_trees}) galaxy are considered, as to avoid combining the SFR histories of multiple progenitors.}. By collectively binning the stellar particles in the galaxy at $z=0$ by their birth time, weighting by their initial mass and dividing by the bin width, a robust SFR history is obtained, which is only limited in resolution by the total number of stellar particles sampled (i.e., the galaxy stellar mass at $z=0$). For galaxies more massive than $M_*[z=0] \geq 10^{10}$ \Msol this can adequately resolve the SFR history down to intervals of $\approx 1$~Myr, which achieves orders of magnitude better sampling than from the snapshot output. This method can be used to accurately study the SFR histories of individual galaxies \citep[for example Figure 1 from][]{McAlpine2017}, or simply as a method to obtain the maximum SFR ever achieved by a galaxy throughout its lifetime.   

\subsubsection{Absolute calibration of SFRs}
\label{sect:sfr_calibration}

When comparing to the observed cosmic SFR density, \citet{Furlong2015} found an almost constant $-$0.2~dex offset for redshifts $z \leq 3$. There is, however, continued uncertainty as to the absolute calibration of SFR indicators on which these observations rely. For example, \citet{Chang2015} find upon revisiting this calibration with the addition of \textit{WISE} photometry to the full SDSS spectroscopic galaxy sample that the SFRs of typical local galaxies are systematically lower than previously estimated by $\approx 0.2$~dex, yielding better agreement with the \eagle prediction \citep[see Figure 5 of][]{Schaller2015b}.

As the observational datasets compared to in \cref{sect:obs_comparison} utilise an earlier calibration, we \textit{reduce} all observed SFRs by 0.2~dex. This serves to remove the known global systematic offset, making it simpler to focus on the trends with the observed \smm population that are the topic of this paper. 

\subsubsection{Mock observables}
\label{sect:radiative_transfer}

The light emitted within a galaxy is subject to attenuation by dust in the interstellar medium (ISM). To accurately produce the observable properties of the model galaxies, we solve the three-dimensional radiative transfer problem in post-processing using information from the galaxy's star-forming regions, stellar sources, and diffuse dust distribution using the radiative transfer code {\sc skirt} \citep{Baes2011,Camps2015}, as is detailed by \citet{Camps2018}. This process infers the mock observables from the UV to the \smm wavebands\footnote{The rest-frame magnitudes and observer-frame fluxes for all \eagle galaxies with stellar masses greater than $10^{8.5}$ \Msol are publicly available \citep{Camps2018}.}. The free parameters in the radiative transfer model\footnote{Notably for the dust distribution: $f_{\mathrm{dust}}$, the fraction of the metallic gas locked up in dust (assumed to be $0.3$); and $T_{\mathrm{max}}$, the highest temperature at which gas contains dust \citep[assumed to be 8000K, see][]{Camps2016}.} have been calibrated to reproduce far-infrared observables from the \emph{Herschel} Reference Survey \citep{Boselli2010} in the local Universe. Given the weaker observational constraints for these parameters at higher redshifts, the values obtained locally are applied to the galaxies at all redshifts, however this is likely to be an oversimplification \citep[see][ for more details]{Camps2016,Trayford2017}.

For this analysis, we compare to the observed \smm population using the {\sc skirt} inferred observed-frame fluxes at 850$\mu$m \citep[\fsmm, column {\sc SCUBA2\_850} in the public database,][]{McAlpine2015,Camps2018}.

\subsection{Sample selection}
\label{sect:sample_selection}

The primary motivation for this study is to infer the nature of the observed SMG population using analogues from the simulation. With the advent of ALMA the properties of the observed \smm population are becoming better constrained, with the bulk of the population found at higher redshifts \citep[$z > 1$, e.g.,][]{Simpson2014}; as are the  majority of highly star-forming sources found through other multi-wavelength selection techniques \cite[$\dot M_* \gtrsim 100$~\Msolyr, e.g.,][]{MadauandDickinson2014}. For simplicity, in this study we therefore concentrate our analysis on the \smm and highly star-forming model galaxies from the simulation above $z \geq 1$, where the vast majority of the observational constraints on these populations exist. We will return to the \smm properties of low-redshift ($z<1$) galaxies in a future study.

\subsubsection{Sample 1: \smmbright galaxies}
\label{sect:sample_selection_smmbright}

We select galaxies with mock \smm fluxes greater than 1~mJy (\fsmm $\geq 1$~mJy). This cut broadly reflects the definitions in the literature for highly star-forming sources derived from the recent ALMA studies \citep[e.g.,][]{Simpson2014,Cowie2018}, and produces a selection of 62 \smmbright model galaxies in \eagle at $z \geq 1$.

\subsubsection{Sample 2: highly star-forming \smmfaint galaxies}
\label{sect:sample_selection_smmfaint}

 The observed \smm population is synonymous with high SFRs ($\dot M_* \gtrsim 100$~\Msolyr), which is also true for the model galaxies in the \smmbright sample (see \cref{sect:SFRs}). Yet it is interesting to ask if the reverse also holds, that is, are all \squotes{highly star-forming} sources bright in the \smm? To test this, we select the most highly star-forming galaxies in the simulation at $z \geq 1$, adopting a space density which is representative of the limits on highly star-forming populations derived from a variety of observational techniques. There is, however, considerable uncertainly in this quantity observationally. For example, surveys which select sources on the basis of their mid- or far-infrared luminosity (as a proxy for their SFR) have reported space densities of $\sim 10^{-5}$ to $2 \times 10^{-4}$ cMpc$^{-3}$ for galaxies with far-infrared luminosities of $\gtrsim 10^{12}$ L$_\odot$ and inferred SFRs of $\gtrsim 100$ \Msolyr at $z \approx$ 1.5--2.5 \citep[e.g.,][]{Chapman2005,Magnelli2011,Casey2012,Gruppioni2013,Swinbank2014,Koprowski2017}. However, these studies suffer from a combination of AGN contamination of the derived luminosities \citep[a particular problem in the mid-infrared, e.g.,][]{Kirkpatrick2012,Kirkpatrick2015,DelMoro2013}, or blending and misidentification of the correct galaxy counterparts in low-resolution far-infrared and sub-millimeter surveys \citep[e.g.,][]{Hodge2013,Simpson2014}. 

Given the range and uncertainties in the various estimates of the space density for highly star-forming galaxies from the observations, we have chosen to select an equivalent SFR limit which roughly corresponds to the space densities derived from the observations ($\sim 10^{-5}$ to $2 \times 10^{-4}$ cMpc$^{-3}$) and which allows us to isolate a sufficiently large sample ($\gtrsim 100$ galaxies) of highly star-forming galaxies to allow for a statistical analysis. Our adopted SFR limit is $\geq 80$ \Msolyr, which corresponds to a space density of star-forming galaxies in \eagle of $8 \times 10^{-5}$ cMpc$^{-3}$ at $z \geq 1$, and returns a sample of 84 galaxies.

We find, perhaps surprisingly, that only 32\% of the galaxies at $z \geq 1$ with a SFR $\geq 80$~\Msolyr have mock \smm fluxes \fsmm $\geq 1$~mJy, implying that a large number of highly star-forming galaxies in the Universe may not be detected by the current \smm surveys. For this reason, we additionally investigate the nature of highly star-forming galaxies ($\dot M_* \geq 80$~\Msolyr) that are \squotes{\smmfaint} (\fsmm $< 1$~mJy), to see how, if at all, these galaxies differ from the \smmbright population. This will also reveal what subset of highly star-forming galaxies is selected by the \smm surveys.

We note, that we have repeated the analysis of this study for highly star-forming \smmbright galaxies (i.e., adding the additional $\dot M_* \geq 80$~\Msolyr criterion to define the \smmbright sample), however the results were almost identical to those found for the \smmbright only sample (as the majority of these galaxies are naturally above $\dot M_* \geq 80$~\Msolyr). We remain with the pure \smm-selected sample for this study as it most closely reflects that of the observations.

\section{Results}

\begin{figure}
\includegraphics[width=\columnwidth]{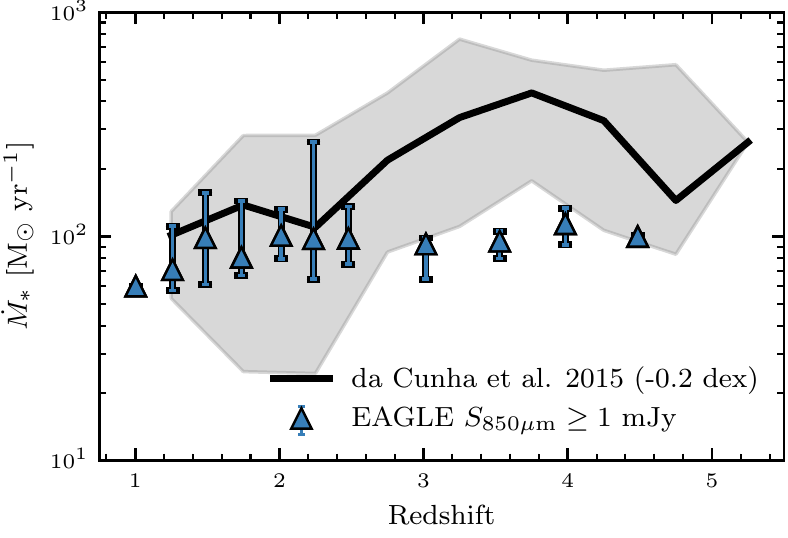}

\caption{The median SFR as a function of redshift for both the \eagle \smmbright galaxies at each simulation snapshot and the observed SMGs from \citet{daCunha2015}. A $-$0.2~dex SFR recalibration has been applied to the observational data at all redshifts (see \cref{sect:sfr_calibration}). The error bars and the height of the shaded region outline the $10^{\mathrm{th}}-90^{\mathrm{th}}$ percentile range. The SFRs of the model galaxies in the \smmbright sample are high ($\dot M_* > 50$~\Msolyr, with a median SFR of 94~\Msolyr), and are in reasonable agreement with the SFRs inferred from the observations below $z \approx 3$ where the bulk of the population lie. However, the SFRs of the model galaxies in the high-redshift tail ($z \gtrsim 3$) are a factor of $\approx 3$ lower than the observations.}

\label{fig:sfr_distribution}
\end{figure}

\begin{figure*}
\includegraphics[width=\textwidth]{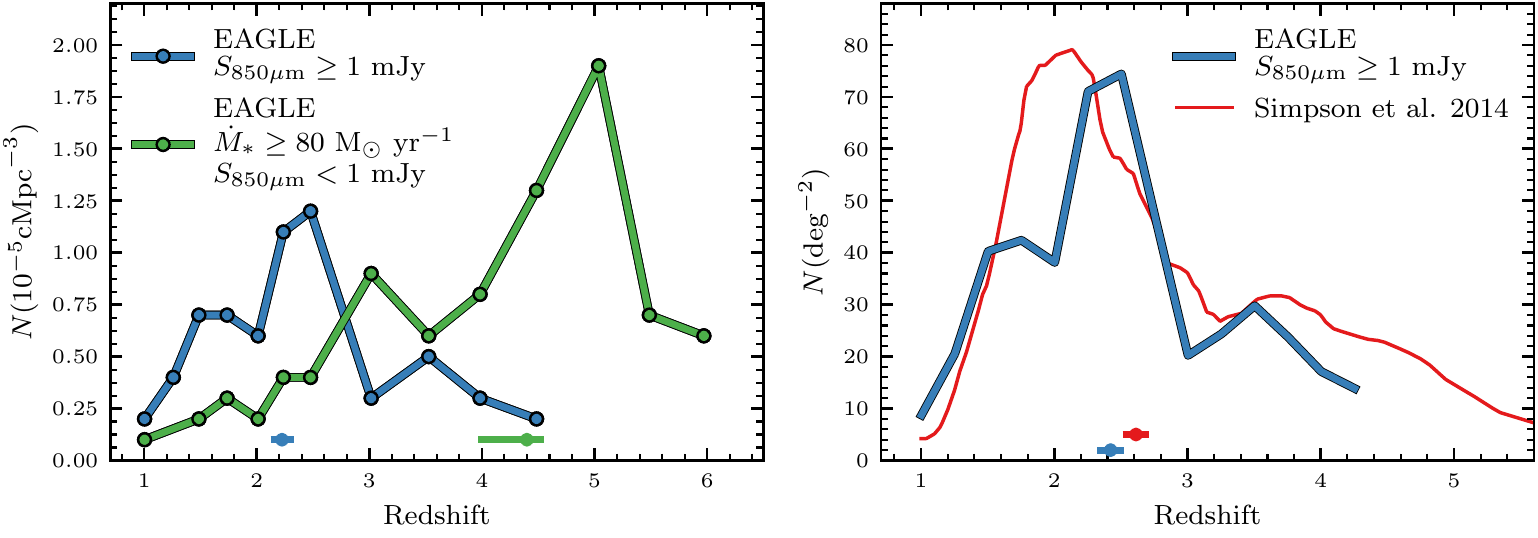}

\caption{\textit{Left panel}: the number density of the \smmbright and highly star-forming \smmfaint model galaxies (in units of 10$^{-5}$cMpc$^{-3}$) at each simulation snapshot. The median redshift of each sample is indicated by an error bar along the lower axis ($z=2.2_{-0.1}^{+0.1}$ and $z=4.4_{-0.4}^{+0.1}$ for the \smmbright and highly star-forming \smmfaint population, respectively). The galaxies in the \smmbright sample are most abundant at $z \approx 2.5$ and then become increasingly rare towards lower and higher redshift. The galaxies within the highly star-forming \smmfaint sample are predominantly at higher redshift ($z > 4$). \textit{Right panel}: the number of \smmbright model galaxies per square degree (median value of $z=2.4_{-0.1}^{+0.1}$) compared to the statistically corrected observed sample of SMGs from \citet{Simpson2014} (median value of $z=2.6_{-0.1}^{+0.1}$). The shape and the median value are similar for the two distributions. The position of the peak is slightly higher for the sample of \smmbright model galaxies ($z \approx 2.5$) compared to the observations ($z \approx 2$).}

\label{fig:redshift_distribution}
\end{figure*}

\begin{figure*}
\includegraphics[width=\textwidth]{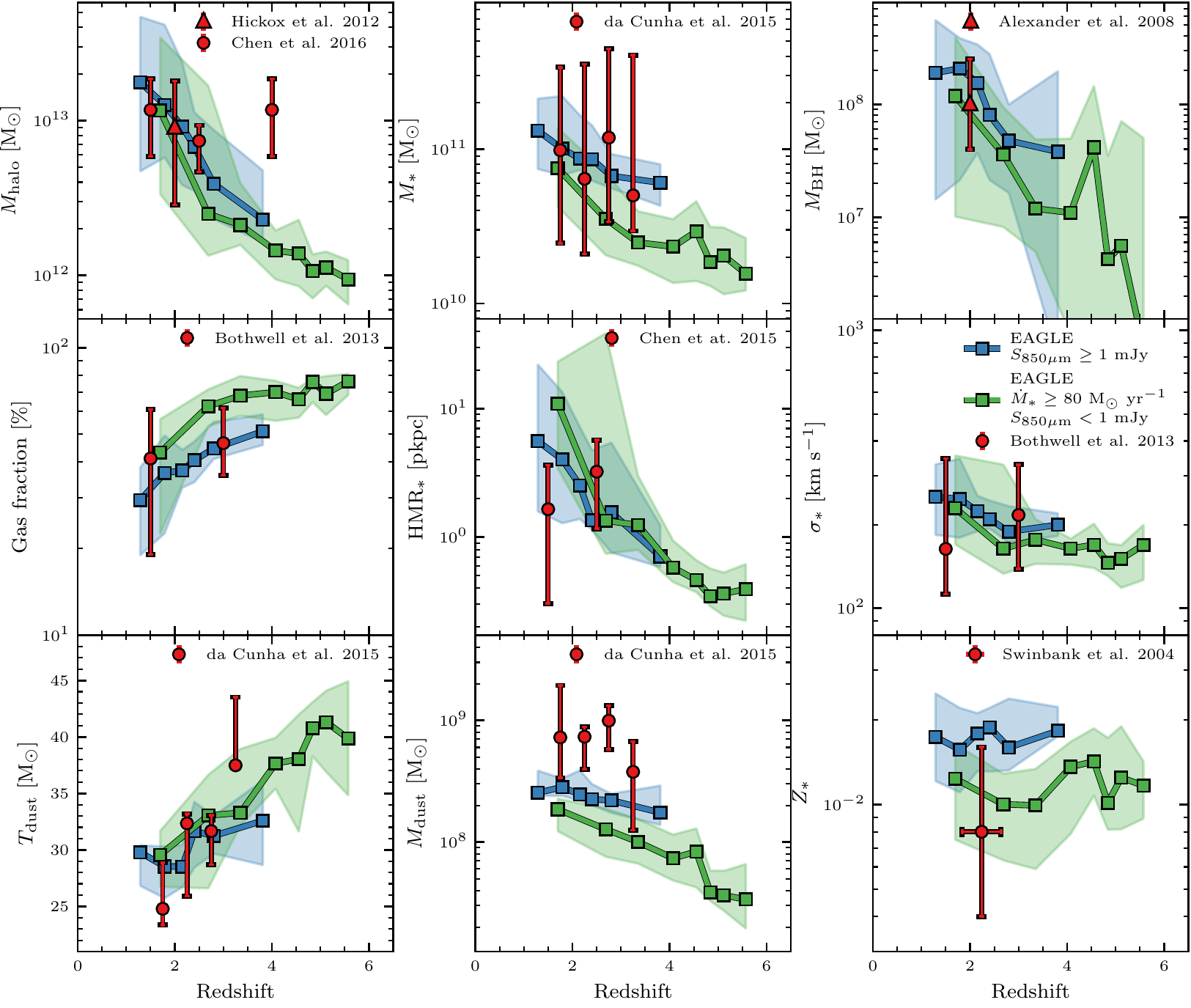}

\caption{Various properties of the model galaxies in the \smmbright (blue) and highly star-forming \smmfaint (green) samples, each plotted as a function of redshift. The median values of the \eagle galaxies within each sample are represented by lines, with the shaded regions outlining the $10^{\mathrm{th}}-90^{\mathrm{th}}$ percentile ranges. Each bin is ensured to contain at least 10 galaxies. The red data points show estimated from observations of SMGs, and should therefore only be compared to the \smmbright sample (shown in blue). \textit{Top left}: the halo mass, with observations from \citet{Hickox2012} and \citet{Chen2016}. \textit{Top center}: the total stellar mass, with observations from \citet{daCunha2015}. \textit{Top right}: the central supermassive black hole mass, with observations from \citet{Alexander2008}. \textit{Middle left}: the total gas fraction, $M_{\mathrm{gas}} / M_{\mathrm{gas+stars}}$, with observations from \citet{Bothwell2013}. \textit{Middle center}: the stellar half mass radius, with observations from \citet{Chen2015}. \textit{Middle right}: the stellar velocity dispersion, with observations from \citet{Bothwell2013}. \textit{Bottom left}: the dust temperature, with observations from \citet{daCunha2015}. \textit{Bottom center}: the dust mass, with observations from \citet{daCunha2015}. \textit{Bottom right}: the stellar metallicity, with observations from \citet{Swinbank2004}. The \smmbright galaxies in the \eagle model are in reasonable agreement with the observed \smm population across a range of observable properties. The \eagle median values from this figure are tabulated in \cref{tab:prop_smm_bright,tab:prop_smm_faint}.}

\label{fig:smg_properties_at_time}
\end{figure*}

\subsection{A comparison of simulated and observed SMGs}
\label{sect:obs_comparison}

We begin with a comparison between the model galaxies in the \smmbright sample and the results from a variety of \smm observations. Here, our aim is to establish if a \smm selection applied to \eagle serves to extract a population similar to that seen in the real Universe.

In the following, we will compare to the observed SMG population focusing on the first large ALMA-selected sample from the ALESS survey \citep{Hodge2013}. Subsequent analysis and study of the SMGs in this survey have provided a variety of empirical constraints on their natures, which we can compare to our model predictions \citep[e.g.,][]{Simpson2014,daCunha2015}. We also make use of the earlier survey of \citet{Chapman2005} when discussing certain aspects, such as gas or black hole properties \citep[e.g.,][]{Alexander2008,Bothwell2013}.

\subsubsection{Star-formation rates}
\label{sect:SFRs}

The SFRs inferred for the observed \smm population are high ($\dot M_* \gtrsim 100$~\Msolyr), which provides a good initial test for our model population. \cref{fig:sfr_distribution} shows the median SFR ($\dot M_*$) as a function of redshift for both the model galaxies in the \smmbright sample and the observed SMGs from \citet{Hodge2013} as analysed by \citet{daCunha2015}. 

Overall, the SFRs of the \smmbright model galaxies are high ($\dot M_* \geq 50$~\Msolyr), yielding a relatively constant median value between $\approx$ 70--100~\Msolyr for all redshifts, and a maximum SFR of 294~\Msolyr produced by a galaxy at $z=2.2$. This confirms that a purely \smm-selected sample of model galaxies does only return high SFR sources. When compared to the SFRs inferred from the observations, there is good agreement at $z \lesssim 3$ where the bulk of the model and observed population lie, however, a discrepancy appears above $z \gtrsim 3$, where the SFRs of the \smmbright model galaxies are a factor of $\approx 3$ lower than those inferred from observations. It is important to stress, however, that most of the observed and simulated \smm galaxies exist below $z=3$ ($\approx 70$\%, see next section), where there is good agreement between the SFRs from the two populations.    

\subsubsection{Redshift distribution}
\label{sect:redshift_distribution}

The spectroscopic redshifts for large surveys of observed SMGs suggest that this population is most abundant at redshift $z \approx 2$--3 \citep[e.g.,][]{Chapman2005,Danielson2017}. However these are incomplete, and so here we compare the redshift distribution of the model galaxies in the \smmbright sample against the statistically corrected sample of observed SMGs using photometric estimates from \citet{Simpson2014}.

 \cref{fig:redshift_distribution} shows the volume number density of the model \smmbright galaxies at each simulation snapshot. It indicates that this population is most abundant at $z \approx 2.5$ (with a number density of $\sim 10^{-5}$cMpc$^{-3}$), and is rarer towards both lower and higher redshift ($\sim 10^{-6}$cMpc$^{-3}$ at redshifts of $\approx 1$ and $\approx 4$). To provide a measurement that is more easily comparable to the observations, \cref{fig:redshift_distribution} also shows the area number density (or counts) of the model \smmbright galaxies. To compare, we over-plot the redshift distribution of observed SMGs from \citet{Simpson2014}. Encouragingly, the model \smmbright galaxies yield a similar shape and median value ($z_{\mathrm{EAGLE}}=2.4^{+0.1}_{-0.1}$ compared to $z_{\mathrm{obs}}=2.6^{+0.1}_{-0.1}$) as the observations. Yet, whilst the overall behaviour between the two samples appears consistent (a peaked distribution that decays towards lower and higher redshift), the location of the peak for the model \smmbright galaxies is slightly higher than is seen by the observations ($z_{\mathrm{EAGLE}} \approx 2.5$ compared to $z_{\mathrm{obs}} \approx 2$)

\subsubsection{Galaxy and halo properties}

\begin{figure*}
\includegraphics[width=\textwidth]{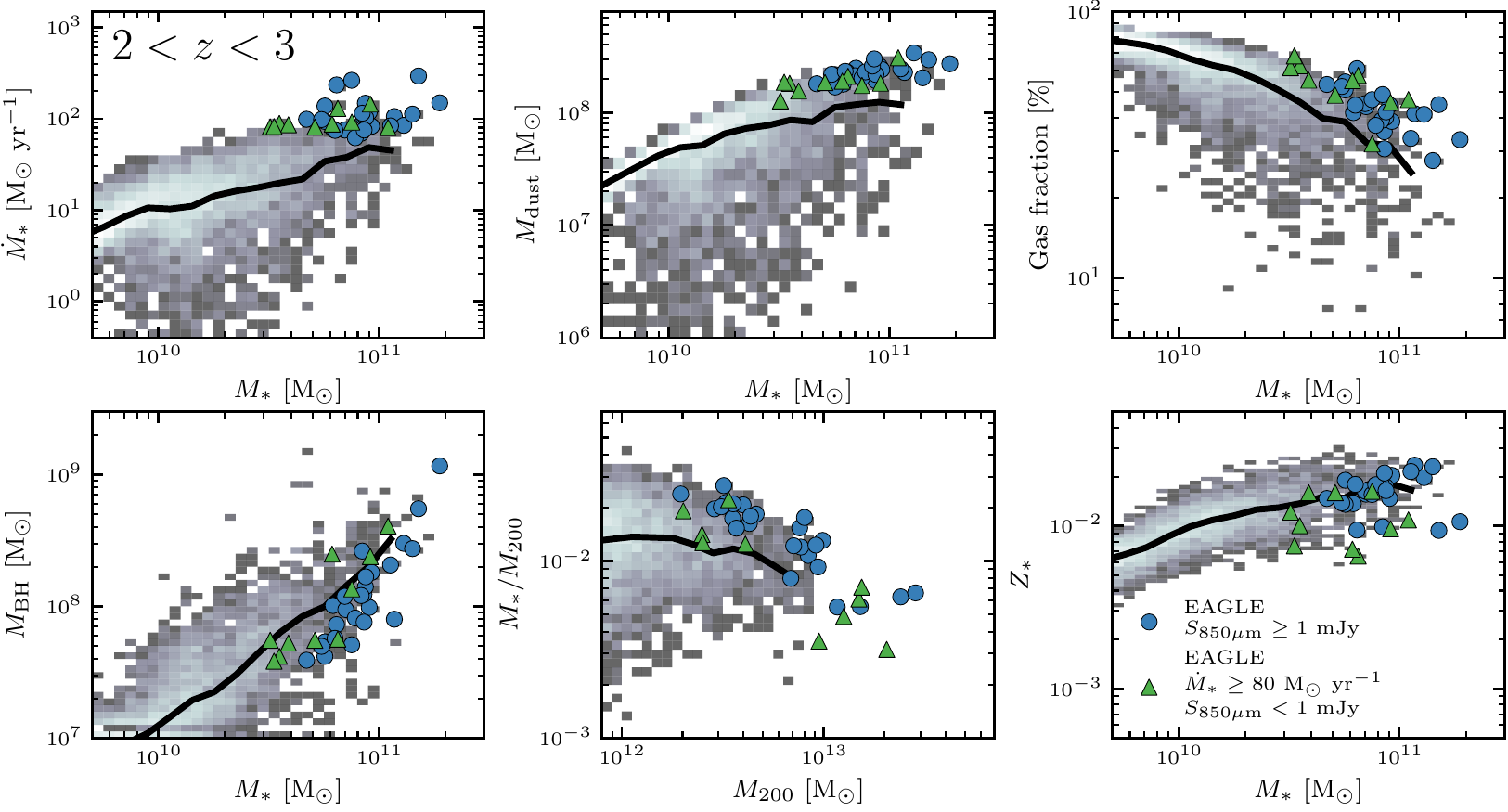}

\caption{The properties at $z = 2$--3 of the model galaxies in the \smmbright and highly star-forming \smmfaint samples (shown individually), plotted as a function of either the stellar or halo mass. We show the SFR (top left), dust mass (top centre), total gas fraction ($M_{\mathrm{gas}} / M_{\mathrm{gas + stars}}$, top right), central suppermassive black hole mass (bottom left), stellar mass to halo mass ratio (bottom centre) and stellar metallicity (bottom right). The general population of all model galaxies in the simulation is represented by a two-dimensional histogram, coloured by the number of galaxies in each bin. The median trend for all model galaxies is indicated via a solid line. Both the \smmbright galaxies and the highly star-forming \smmfaint galaxies are typically massive ($M_* \approx 4 \times 10^{10}$--$2 \times 10^{11}$~\Msol), have high SFRs (by construction) and have the highest dust masses and gas fractions for their stellar masses, all relative to the median trends of the general population in the simulation. In addition, the galaxies within each sample show evidence of hosting undermassive black holes for their stellar masses and the \smmbright galaxies have stellar masses that are high for their halo masses.}

\label{fig:general_population}
\end{figure*}

 Next, in \cref{fig:smg_properties_at_time} we compare a variety of properties of the model galaxies in the \smmbright sample against a the observed population. Each property is plotted as a function of redshift, and shows: the halo mass ($M_{\mathrm{halo}}$), the stellar mass ($M_{\mathrm{*}}$), the central supermassive black hole mass ($M_{\mathrm{BH}}$), the total gas fraction ($M_{\mathrm{gas}} / M_{\mathrm{gas + stars}}$), the stellar half-mass radius (HMR$_{\mathrm{*}}$), the stellar velocity dispersion\footnote{The one dimensional velocity dispersion of the stars, see Table B.1 from \citet{McAlpine2017}.} ($\sigma_{\mathrm{*}}$), the dust mass ($M_{\mathrm{dust}}$), the dust temperature ($T_{\mathrm{dust}}$) and the stellar metallicity ($Z_{\mathrm{*}}$). The values and associated errors for the properties of the model galaxies quoted below are the median and the 1$\sigma$ uncertainties on the median, respectively (these values are tabulated in \cref{tab:prop_smm_bright}).

The model \smmbright galaxies reside in halos of mass $(9.1_{-0.2}^{+4.8}) \times 10^{12}$ \Msol, have stellar masses of ($8.7_{-2.6}^{+0.1}) \times 10^{10}$ \Msol and host black holes of mass $(15.4_{-1.4}^{+9.6}) \times 10^{7}$ \Msol at redshift $z \approx 2$. These values agree well with the observed clustering halo mass estimates from \citet{Hickox2012} and \citet{Chen2016}, the stellar mass estimates from \citet{daCunha2015} and the black hole mass estimates from \citet{Alexander2008}, respectively. The total gas fractions ($37.4_{-1.7}^{+3.2}$\% at $z \approx 2$), stellar sizes ($2.5_{-0.5}^{+0.4}$~pkpc at $z \approx 2$) and velocity dispersions ($223_{-17}^{+25}$~km~s$^{-1}$ at $z \approx 2$) yield reasonable agreements for a range of redshifts to the observations from \citet{Bothwell2013}, \citet{Chen2015} and \citet{Bothwell2013}, respectively. The dust temperatures ($38.5_{-0.9}^{+0.4}$~K at $z \approx 2$) match well to the observations from \citet{daCunha2015} for a range of redshifts, however there is a systematic discrepancy at all redshifts between the dust masses \citep[again from][]{daCunha2015} predicted by the model (($2.5_{-0.5}^{+0.2}) \times 10^{8}$ \Msol at $z \approx 2$) and the observations, with the observed \smm galaxies containing $\approx$ 3--4 times more dust than the model galaxies. Finally, the metallicities ($17.9_{-1.8}^{+0.1} \times 10^{-3}$ at $z \approx 2$) are potentially up to a factor of $\approx 2$ higher than those estimated by \citet{Swinbank2004}, but are consistent to within the errorbars.

Although the observational uncertainties are typically large (often larger than the predicted scatter from the model), the model galaxies within the \smmbright sample are in reasonable agreement with a variety of measurements from a variety of observations of \smm-selected galaxies.

\subsection{The nature of \smmbright galaxies and highly star-forming \smmfaint galaxies}
\label{sect:nature_of_smm}

Given the broad agreement in the comparisons to the observed \smm population above, we are encouraged to exploit the sample of \smmbright model galaxies in an attempt to answer the questions that the observations cannot easily address; such as the origin, evolution and eventual fate of the \smm population. In a parallel analysis, we investigate the nature of the highly star-forming galaxies that do not make it into the \smmbright sample (i.e., $\dot M_* \geq 80$~\Msolyr but \fsmm $< 1$~mJy), to better understand what subset of high-SFR galaxies a \smm selection extracts.

It is important to remind the reader for this section, that the model galaxies in the \smmbright sample are substantially more abundant than the model galaxies in the highly star-forming \smmfaint sample at intermediate redshift ($z \approx 2.5$), whereas the galaxies in the highly star-forming \smmfaint sample are substantially more abundant than the galaxies in the \smmbright sample at higher redshift ($z > 4$, see \cref{fig:redshift_distribution}). Therefore, whilst we can compare the properties of the two model populations at a given redshift, the relative number density of galaxies between the two populations at the given redshift may be considerably different.

\subsubsection{Comparison to the general population}
\label{sect:general_population}

We first investigate how the properties of the model galaxies in our two samples compare to the general model population. That is, what unique features do the \smmbright and highly star-forming \smmfaint galaxies exhibit that distinguish them from \squotes{typical} model galaxies of their mass? 

\cref{fig:general_population} shows the SFR, dust mass, total gas fraction, central supermassive black hole mass, stellar mass to halo mass ratio ($M_* / M_{\mathrm{200}}$) and the stellar metallicity of the galaxies in the \smmbright and highly star-forming \smmfaint samples, each plotted as a function of either the stellar or halo mass. For clarity, we only show the model galaxies in the redshift range $z=2$--3, as these epochs contain the greatest number of sources between our two samples (see \cref{fig:redshift_distribution}). We note, that only the integrated properties of central galaxies are shown in this figure (the galaxies contained within both the \smmbright and highly star-forming \smmfaint samples are almost exclusively centrals at these times).  

Upon first inspection, it is clear that the galaxies across both samples are among the most massive in the model universe, yet, as was indicated previously in \cref{fig:smg_properties_at_time}, there is a preference for the \smmbright galaxies to be more massive (a median stellar mass of $8.4 \times 10^{10}$~\Msol for the \smmbright sample compared to $5.6 \times 10^{10}$~\Msol for the highly star-forming \smmfaint sample). By construction, each sample is probing the galaxies with the highest SFRs for their stellar mass, all lying well above the median trend, which is inextricably linked to their high dust masses and gas fractions. The two model populations also show evidence of hosting black holes that are undermassive for galaxies of their stellar masses, and the \smmbright model galaxies are high in their stellar masses for their halo masses. Finally, the stellar metallicities of the model galaxies within each sample appear consistent with the median trend of the general population, although there is a hint that the galaxies from the highly star-forming \smmfaint sample have metallicities lower than expected for their stellar mass. 

Therefore the model galaxies in both the \smmbright and highly star-forming \smmfaint samples do not appear as \squotes{typical} star-forming galaxies in the simulation; they are massive, unusually gas and dust rich galaxies that host undermassive black holes. A simple explanation is that the two model populations comprise the most massive galaxies who, for one reason or another, host an undermassive black hole. This is consistent with the results presented by \citet{Matthee2018}, who find that the relative efficiently of black hole growth contributes to the scatter in the high-mass end of the star-forming sequence in the \eagle simulation. Such an undermassive black hole would allow the model galaxies to retain a greater amount of gas in their centres due to a decreased cumulative amount of AGN feedback. In addition, having stellar masses which are high for their halo masses suggests that the \smmbright model galaxies are undergoing a somewhat prolonged period of high SFRs, whereas the highly star-forming \smmfaint population have potentially initiated their starburst much more recently. This would be consistent with the highly star-forming \smmfaint galaxies having lower stellar metallicities. 

\begin{figure*}
\includegraphics[width=\textwidth]{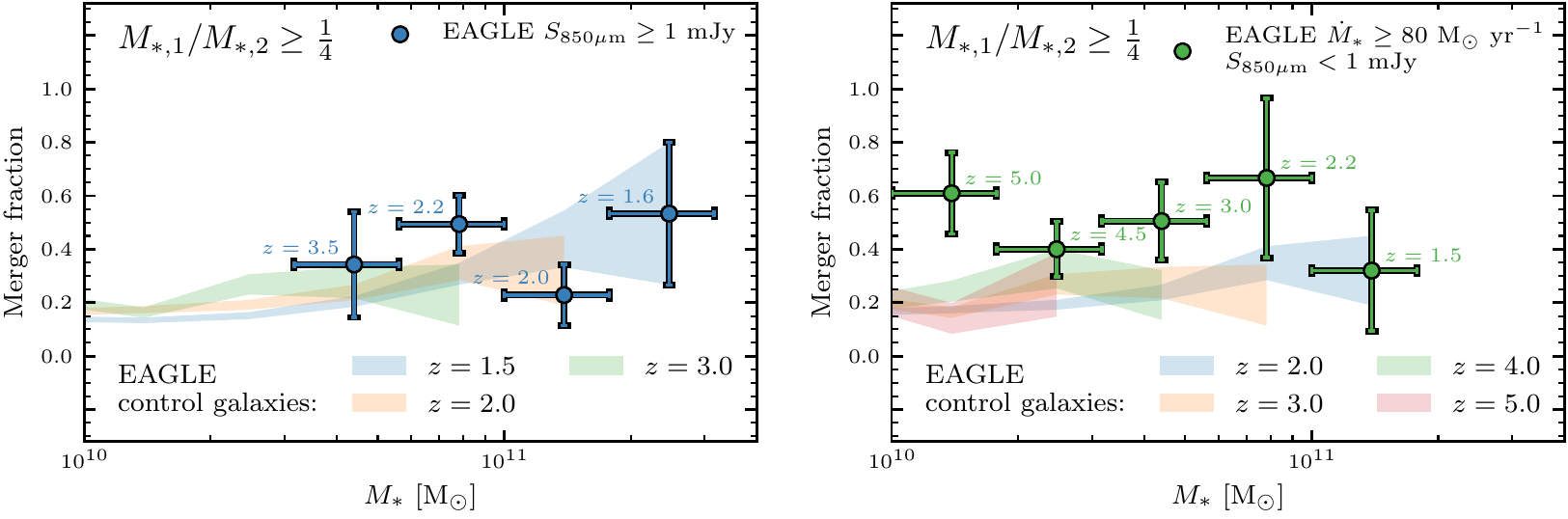}

\caption{The major merger fraction for the two model samples, defined as the fraction of galaxies that have completed or will complete a major merger (\mergerratio $\geq \frac{1}{4}$) within the previous/next dynamical time, as a function of stellar mass. The merger fractions are shown separately for the model \smmbright galaxies (left) and the model highly star-forming \smmfaint galaxies (right), with the merger fractions of all the galaxies in the simulation at a given redshift shown as shaded regions, acting as our control. The vertical errors in the merger fractions and the height of the shaded regions represent the Poisson uncertainty. The errors on $M_*$  indicate the extent of each stellar mass bin. Each datapoint is annotated with the median redshift of the galaxies in the given stellar mass bin. The merger fractions of the \smmbright galaxies are consistent with the general model population, suggesting major mergers are not the primary trigger of the model SMG population. The highly star-forming \smmfaint galaxies have merger fractions typically above the general population, (with the greatest excess coming at lower stellar masses $M_* \ll 10^{11}$~\Msol, where it is a factor of $\approx$ 3), suggesting major mergers are the main driver for triggering this galaxy subset.}

\label{fig:merger_fraction}
\end{figure*}

\subsubsection{The evolution in the integrated properties for the \smmbright and highly star-forming \smmfaint galaxies}

Relative to the the median trend of the general population, the \smmbright and highly star-forming \smmfaint galaxies always have high SFRs, dust masses and gas fractions for their stellar masses (see \cref{fig:general_population}). However, the typical values of the integrated properties for the \smmbright and highly star-forming \smmfaint galaxies do evolve as a function of redshift, as was shown in \cref{fig:smg_properties_at_time}. 

In general, the \smmbright galaxies at higher redshift are less massive, have higher gas fractions, are more compact and have higher dust temperatures relative to their lower-redshift counterparts. These trends also broadly represent the evolution of the properties of the galaxies in the highly star-forming \smmfaint sample. Yet, although the trends are similar, the median values of the two populations are commonly offset from one another at a given redshift. For example; the halo, stellar and black hole masses of the highly star-forming \smmfaint galaxies are a factor of $\approx$ 2--4 times lower than the \smmbright galaxies (see \cref{fig:smg_properties_at_time}), they are additionally more gas rich and more metal poor (both by up to a factor of $\approx 2$), their dust temperatures are a few degrees higher, and their dust masses are a factor of $\approx 2$ lower.

It appears, therefore, that the model galaxies that make up the two populations are different in several aspects, potentially suggesting two alternate formation processes. The galaxies from the highly star-forming \smmfaint subset are, on average; lower mass, more gas rich, more metal poor, contain less and slightly warmer dust, host lower mass black holes and preferentially exist at higher redshifts (see \cref{fig:redshift_distribution}) than the galaxies from the \smmbright subset. Regardless of their classification, however, the galaxies from both samples are always the most gas rich and typically the most massive at their respective redshifts. 

\subsubsection{Merger fractions}
\label{sect:merger_fractions}

Galaxy--galaxy mergers and interactions provide one potential triggering mechanism for starbursting galaxies: the induced tidal field disturbs any regular orbits of the gas, funnels material inward towards the galaxy centre, and triggers star formation \citep[e.g.,][]{BarnesAndHernquist1991,MihosAndHernquist1996}. It is reasonable to assume, therefore, that such interactions may be responsible for creating the observed \smm and highly star-forming galaxy populations, which we test here.

In \cref{fig:merger_fraction} we investigate the major merger fraction, defined as the fraction of galaxies that have completed or will complete a major merger (\mergerratio $\geq \frac{1}{4}$) within the previous/next dynamical time\footnote{Equivalent to $|n_{\mathrm{dyn}}| \leq 1$ from eq. 1 in \citet{McAlpine2018}, where $n_{\mathrm{dyn}}$ is the number of dynamical times to the nearest galaxy--galaxy merger with a stellar mass ratio $\geq \frac{1}{4}$. The dynamical time, $t_{\mathrm{dyn}}$, is defined as the free fall time of the dark matter halo ($t_{\mathrm{dyn}} \approx$ 1.6~Gyr at $z=0$, $\approx$ 0.5~Gyr at $z=2$ and $\approx$ 0.2~Gyr at $z=5$), see Section 2.3 from \citet{McAlpine2018}.}, as a function of the stellar mass. The major merger fractions are shown separately for the model \smmbright galaxies and the model highly star-forming \smmfaint galaxies, with the merger fractions for all model galaxies in the simulation at a given redshift shown as shaded regions, acting as our control. Due to their limited size, we only divide the two samples into bins of stellar mass and not also by redshift, realising that the merger fraction of galaxies for a fixed stellar mass increases with increasing redshift \citep[e.g.,][]{Rodriguez-Gomez2015,Qu2017}. However, this effect is accounted for in part by expressing the merger fraction using a set number of dynamical times instead of using a fixed time interval. In addition, separating the galaxies by their stellar masses also acts to separate them by their redshift, as is indicated by the highlighted median redshift of the galaxies in each stellar mass bin (see also \cref{fig:smg_properties_at_time}). Yet, there could remain a further dependence on the importance of mergers in creating the \smmbright/highly star-forming \smmfaint populations with redshift \citep[potentially similar to the discovered redshift dependence on mergers for triggering the rapid growth phase of black hole growth in the \eagle simulation, see][]{McAlpine2018}.

\begin{figure*}
\includegraphics[width=\textwidth]{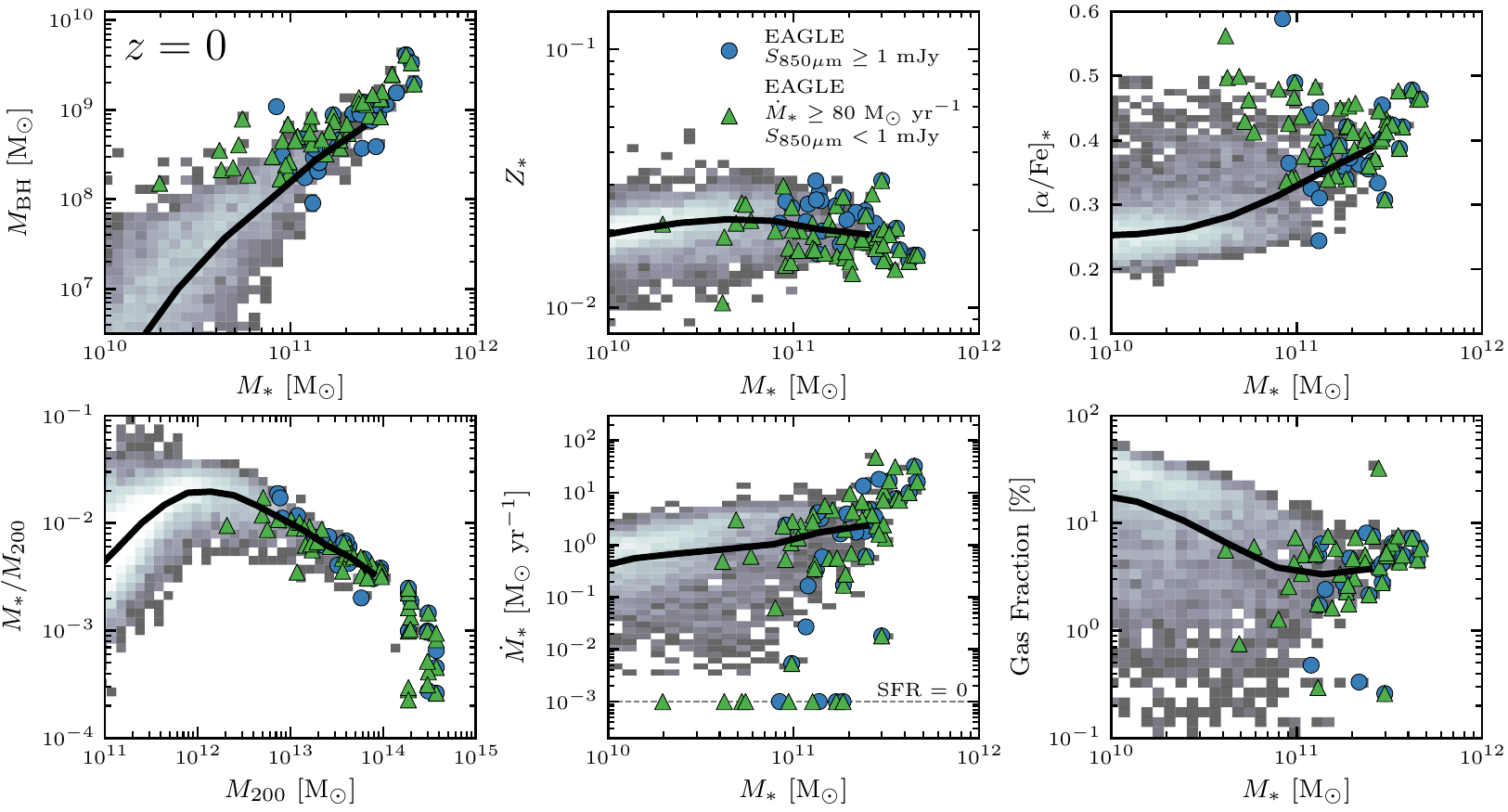}

\caption{The properties at $z=0$ of the descendants of the model galaxies in the \smmbright and highly star-forming \smmfaint samples plotted as a function of either the stellar or halo mass. We plot the central suppermassive black hole mass (top left), stellar metallicity (top centre), alpha-enhancement (top right), stellar mass to halo mass ratio (bottom left), SFR (bottom centre) and total gas fraction ($M_{\mathrm{gas}} / M_{\mathrm{gas + stars}}$, bottom right). The total model galaxy population is represented as a two-dimensional histogram, coloured by the number of galaxies in each bin. The median trend for all galaxies is indicated via a solid line. We assign galaxies with a zero SFR to a value of $\dot M_* = 10^{-3}$~\Msolyr (indicated by a horizontal dashed line on the figure). In all panels, both central and satellite galaxies are shown (except in the lower left panel, where the general population only shows central galaxies for clarity). The once \smmbright galaxies are not obviously distinguishable from the general population at a given stellar mass in any of the integrated properties we have explored (i.e., they follow the median trends). However, the once highly star-forming \smmfaint galaxies have, on average, overmassive black holes, lower metallicities and are more $\alpha$-enhanced for their stellar masses at $z=0$.}

\label{fig:descendants}
\end{figure*}

We find no strong trend in the major merger fraction with stellar mass for either the \smmbright or highly star-forming \smmfaint samples. The major merger fractions of the \smmbright galaxies are consistent with those of the general model population, suggesting that major mergers are not a required triggering mechanism for the model SMGs. What appears to be more important, therefore, is that they have a large gas reservoir and host an undermassive black hole (see \cref{sect:general_population}). By contrast, the major merger fractions of the highly star-forming \smmfaint galaxies are typically higher than those of the general population (particularly at the lowest stellar masses, $M_* \sim 10^{10}$~\Msol, where the enhancement is a factor of $\approx 3$), suggesting that major mergers are important in triggering this galaxy subset.

We note, that whilst the merger fractions universally increase in the case of minor mergers ($\frac{1}{10} \leq$ \mergerratio $< \frac{1}{4}$) and either major or minor mergers (\mergerratio $\geq \frac{1}{10}$), the merger fractions of the \smmbright galaxies are always similar to those of the general population. An enhancement in the merger fraction above the control galaxies for the highly star-forming \smmfaint population exists for all merger classifications, however the enhancement is greatest in the case of major mergers.

\subsection{The descendants of \smmbright galaxies and highly star-forming \smmfaint galaxies at $z=0$}
\label{sect:descendants}

We conclude our analyses by investigating the descendants of the galaxies in both samples at $z=0$. Here we are asking the following: What fraction of today's galaxies have undergone a \smmbright or highly star-forming \smmfaint phase in their past?  Do the galaxies in either of the two samples retain any signatures that could allow them to be identified in today's parameter space? What fraction of the total in situ mass formed within the galaxies from the two samples was formed during high SFR events? 

\cref{fig:descendants} shows the $z=0$ central supermassive black hole mass, stellar metallicity, alpha-enhancement ([$\alpha$/Fe]$_*$\footnote{We define alpha-enhancement following eq. 1 from \citet{Segers2016}, where we use [O/Fe]$_*$ as a proxy for [$\alpha$/Fe]$_*$.}), stellar mass to halo mass ratio, SFR and total gas fraction, each plotted as a function of either the stellar or halo mass. The general population of model galaxies is shown as a two-dimensional histogram and the descendants of the \smmbright and highly star-forming \smmfaint model galaxies are highlighted individually.  

\subsubsection{The fraction of today's galaxies that were once highly star forming}

\begin{figure}
\includegraphics[width=\columnwidth]{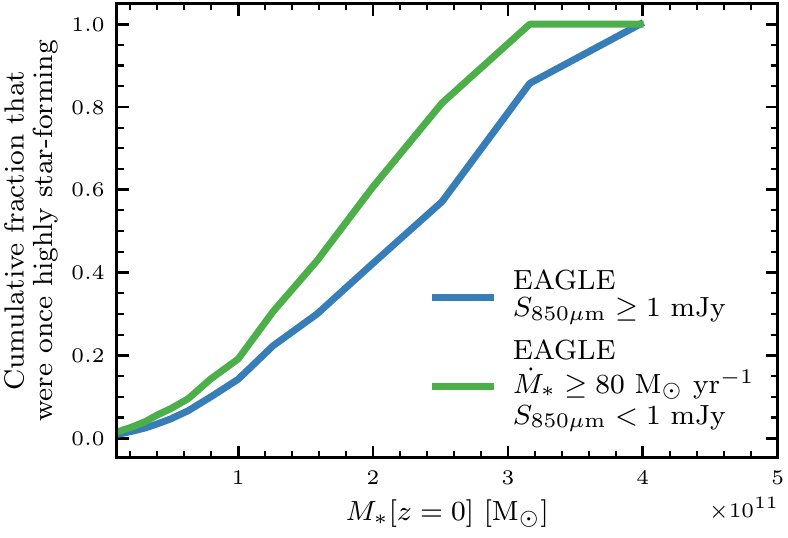}

\caption{The cumulative fraction of galaxies at $z=0$ that were ever once \smmbright or highly star-forming and \smmfaint, as a function of the present day stellar mass. The descendants of these galaxies are rare; only $\approx 20$\% of the galaxies at $z=0$ with stellar masses greater than $M_*[z=0] \geq 10^{11}$~\Msol were once highly star forming. However, this fraction rapidly increases to 100\% for stellar masses $M_*[z=0] \geq 3 \times 10^{11}$~\Msol.}

\label{fig:cumulative_fraction}
\end{figure}

The clearest distinguishing feature seen in the descendants from \cref{fig:descendants}, which was also true at the times when they were selected (see \cref{fig:general_population}), is that they are typically massive (with median masses of $M_* = 2.1 \times 10^{11}$~\Msol and $M_{\mathrm{200}} = 6.9 \times 10^{13}$~\Msol and $M_* = 1.8 \times 10^{11}$~\Msol and $M_{\mathrm{200}} = 6.5 \times 10^{13}$~\Msol for the \smmbright and highly star-forming \smmfaint samples, respectively). Yet, not all massive galaxies today were once highly star forming. To quantify this more clearly, we introduce \cref{fig:cumulative_fraction}, which shows the cumulative fraction of the galaxies at $z=0$, as a function of the $z=0$ stellar mass, that were ever either \smmbright or highly star-forming and \smmfaint in their past (i.e., the cumulative fraction of today's population that are the descendants of the galaxies within either sample). We see a rapid decline in the cumulative fraction for both classifications as the stellar mass decreases; from 100\% of galaxies above $3 \times 10^{11}$~\Msol to just $\approx 20$\% of the galaxies above $1 \times 10^{11}$~\Msol\footnote{We note that a given galaxy at $z=0$ could have been \emph{both} \smmbright and highly star-forming and \smmfaint at different points in its history. Thus the sum of the cumulative fractions between the two samples can exceed 1.}. We note that these fractions serve as a lower limit, as it is unlikely that all instances of galaxies in a high SFR phase can be captured by the temporal spacing's of the simulation snapshots. However, by considering the galaxies today that were ever once highly star-forming (i.e., if they were ever above $\dot M_* \geq 80$~\Msolyr in their past, obtained by examining accurate SFR histories, see \cref{sect:sfr_from_particles}), we find the temporal spacing of the snapshots is sufficient to capture $>80$\% of high SFRs events, making for an almost complete sample.

Therefore, with the exception of the galaxies with the highest stellar masses ($M_* \gtrsim 1$--$3 \times 10^{11}$ \Msol), it is rare for a given galaxy today to of been either \smmbright or highly star-forming and \smmfaint in its past, particularly at lower stellar masses ($M_* < 10^{11}$ \Msol).

\subsubsection{The descendants in today's parameter space}
\label{sect:descendant_properties}

Now we ask what characteristics (at a fixed mass) may be imprinted upon the descendants of the \smmbright and highly star-forming \smm faint galaxies that could potentially distinguish them from the general population of galaxies at $z=0$.

For many of the integrated properties shown in \cref{fig:descendants}, and for the majority of the integrated properties we have explored, we do not find that the descendants of the galaxies in the \smmbright and highly star-forming \smmfaint samples are distinguishable from the general population at $z=0$ (that is, they lie on or around the median trend for galaxies of their stellar/halo masses). For example, they have the expected stellar masses for their halo masses, and the expected sizes, SFRs and total gas fractions for their stellar masses. In fact, for the \smmbright population, no property appears obviously offset away from the median trends of the general population at a given stellar/halo mass. However, some properties do begin to isolate the descendants of the highly star-forming \smmfaint sample. They typically host black holes overmassive for their stellar masses, with the most extreme examples occurring at lower stellar masses ($M_* \ll 10^{11}$~\Msol), suggesting a strong link between high-redshift starbursting galaxies and black hole growth. In addition, their descendants are typically metal poor and alpha-enhanced, the combination of which would suggest they have undergone a rapid stellar mass build up followed by a rapid quenching of continued star formation in their past, which again likely points towards a period of rapid black hole growth \citep[and its associated AGN feedback, see also][]{Segers2016}.

Whilst only a minority of the galaxies in the two samples were classified as satellites at the times that they were highly star-forming (10\% of the \smmbright galaxies and 3\% of the highly star-forming \smmfaint galaxies), 35\% of the \smmbright galaxies and 32\% of the highly star-forming \smmfaint galaxies evolve to become satellites by $z=0$. This would suggest a link between the galaxies within the two samples and their environment. In fact, many of the descendants (particularly from the highly star-forming \smmfaint sample) reside as satellites in today's most massive haloes. For example, in the three most massive haloes from the simulation at $z=0$, 25 of their galaxies were either \smmbright or highly star-forming and \smmfaint in their past. The act of galaxies in-falling into these massive haloes is the cause for a number of the descendants to have \squotes{low} stellar masses (due to stellar stripping), zero SFRs (due to ram-pressure-stripping) and excessively overmassive black holes for their stellar mass \citep[again as a result of the stellar stripping, see][ for a study of these sources]{Barber2016}.  

\subsubsection{The contribution to the in situ stellar mass budget from high SFR events}

\begin{figure}
\includegraphics[width=\columnwidth]{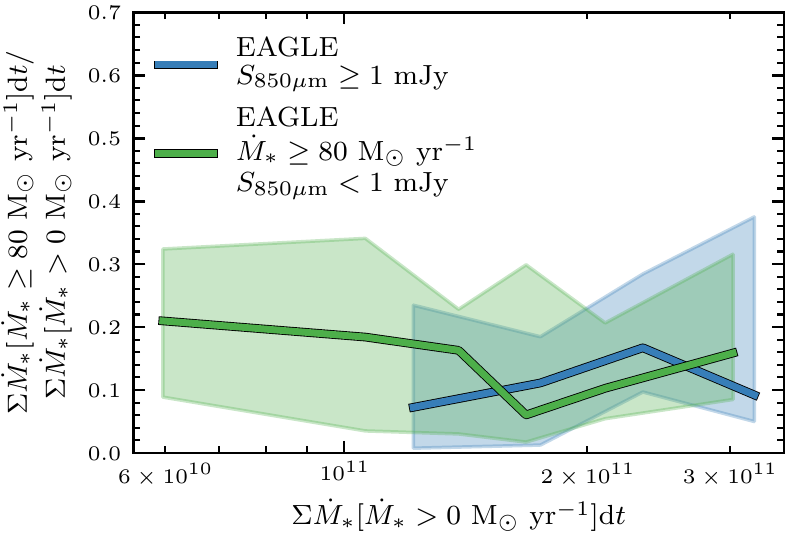}

\caption{For the descendants of the model galaxies in the \smmbright and highly star-forming \smmfaint samples at $z=0$, the fraction of the total initial stellar mass formed in situ within the galaxy at a rate above $\dot M_* \geq 80$~\Msolyr as a function of the total initial mass that formed in situ within the galaxy. On average, the fractions in both samples are low (10--20\%, implying a typical starbursting duration of $\approx$ 100--400~Myr), indicating that high-SFR events are not the main contributor to the in situ production of stars for the galaxies in both samples. Instead, the majority of stars produced in situ within these galaxies are formed at lower SFRs, over longer periods of time.}

\label{fig:mass_in_burst}
\end{figure}

For a galaxy to be in one of the two samples it must achieve an intrinsically high SFR or \smm flux (which also indicates a high SFR, see \cref{fig:sfr_distribution}). We are therefore capturing galaxies at a moment in time when they are among the most highly star-forming objects at their respective redshifts. However, without knowledge of the typical duration of the highly star-forming phase, it is unclear how much this star forming episode will contribute to the total in situ stellar mass budget of the galaxy. That is, is such a phase typically rapid and short lived, and so may contribute relatively little to the total in situ stellar mass budget, or does a significant fraction of the total stellar mass that is grown within the galaxy form during this phase?

\cref{fig:mass_in_burst} shows, for the descendants of both the \smmbright and highly star-forming \smmfaint galaxies at $z=0$, the fraction of the total \emph{initial} stellar mass that was \emph{formed in situ within the galaxy} at a rate above $\dot M_* \geq 80$~\Msolyr, as a function of the total initial stellar mass that was formed in situ within the galaxy (both of these values are computed by integrating the galaxy's entire star formation history, see \cref{sect:sfr_from_particles}). For both samples, periods of high SFRs are never the main contributor to the production of stars within these galaxies, typically contributing an average of 10--20\% to the total in situ initial stellar mass. To put this into context, for an average galaxy that has formed $10^{11}$~\Msol worth of in situ stars by $z=0$, $\approx 20$\% of these were formed during periods of high SFRs, implying a typical duration of $\approx 250$~Myr for the high SFR phase(s) (combined between one or multiple events). Therefore the majority of the in situ stellar mass formed within the descendants of the \smmbright and highly star-forming \smmfaint galaxies occurred at lower SFRs, and over longer periods of time.    

\subsection{The hidden population of highly star-forming galaxies with faint \smm fluxes} 
\label{sect:discussion_hot}

\begin{figure}
\includegraphics[width=\columnwidth]{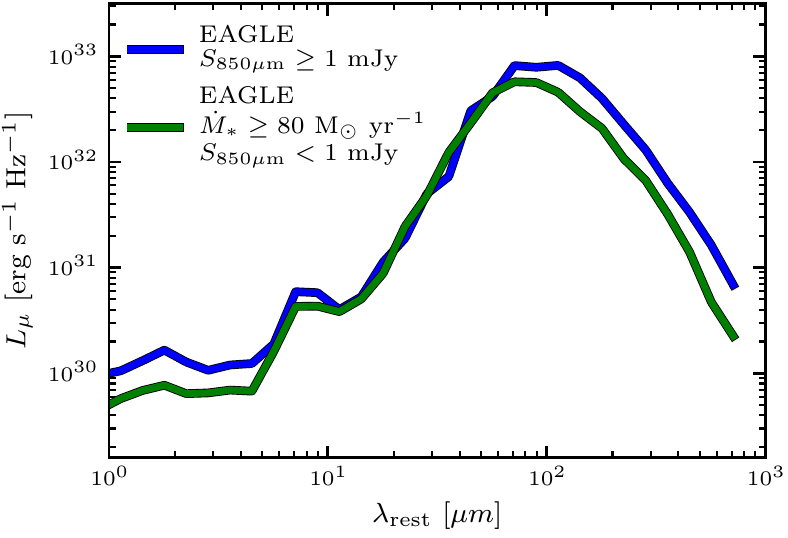}

\caption{Composite broadband spectral energy distributions (SEDs) for the \smmbright and highly star-forming \smmfaint galaxies. The leftward shifting of the peak in the dust emission for the highly star-forming \smmfaint sample signifies hotter dust temperatures than the \smmbright galaxies.}

\label{fig:SED}
\end{figure}

There is increasing empirical evidence that SMGs at higher redshifts ($z \geq 3$) have higher dust temperatures than their lower-redshift ($z \approx 2$) counterparts \citep{Cooke2018,Schreiber2018}. An increasing dust temperature with increasing redshift is also seen in the model galaxies from the \smmbright sample ($29.8_{-0.1}^{+1.5}$~K at $z \approx 1$ to $32.6^{+1.5}_{-0.8}$~K at $z \approx 4$, see \cref{fig:smg_properties_at_time}). This could be linked to physical differences in their galaxy properties, such as their SFR surface densities, which would be consistent with the trend of an approximately constant SFR (see \cref{fig:sfr_distribution}) combined with a decreasing galaxy size with increasing redshift (see \cref{fig:smg_properties_at_time}), as found for the \smmbright model galaxies. However, the model galaxies from the highly star-forming \smmfaint sample have dust temperatures that are systematically higher than those from the \smmbright sample at all redshifts ($29.6_{-1.0}^{+1.7}$~K at $z \approx 2$ to $39.9^{+1.2}_{-1.4}$~K at $z \approx 6$, see \cref{fig:smg_properties_at_time}), which suggests that this may be the reason that they do not make it into the \smmbright sample and would potentially exclude them from current \smm surveys.

The reason an increased dust temperature may exclude the highly star-forming \smm galaxies from the \smmbright sample becomes clear when looking at the spectral energy distributions (SEDs) of the galaxies. \cref{fig:SED} shows the composite broadband SED for the two populations of model galaxies, which clearly reveals a shift to higher temperatures in the dust emission peak for the highly star-forming \smmfaint galaxies relative to the \smmbright galaxies, confirming the hotter dust component, and a corresponding reduction in the the \smm brightness.

The existence of a population of \squotes{hot}, dusty ultraluminous galaxies at high redshift has been suggested previously in the literature \citep[e.g.,][]{Chapman2004,Blain2004,Casey2009}. These galaxies exhibit similar radio and optical characteristics to the high-redshift \smm-selected population, however, they are faint in the \smm, suggesting a hidden population with higher characteristic dust temperatures than the detected \smm population. However, conclusive evidence of a high-redshift population of strongly star-forming \smm faint galaxies, as found by \eagle, has so far remained elusive.

\section{Summary \& Conclusions}
\label{sect:conclusions}

In this study, we have investigated the nature of the model galaxies in the \eagle simulation with the highest SFRs throughout cosmic time. We did this using two model galaxy samples. The first sample comprised the simulated galaxies that yielded the highest mock \smm fluxes (\fsmm $\geq 1$~mJy), which acted as an analogue to the observed \smm population. The second sample contained the simulated galaxies that were defined as highly star-forming ($\dot M_* \geq 80$~\Msolyr), but did not make it into the \smmbright sample (i.e., they are also \fsmm $<1$~mJy, or \squotes{\smmfaint}). This second sample revealed what subset of the model galaxies with high SFRs the \smmbright sample is (or more importantly is not) selecting. 

In \cref{sect:obs_comparison} we began with a comparison to the observed \smm population, finding some encouraging similarities when compared against the \smmbright model galaxy sample. First, a purely \smm-based selection returned the simulated galaxies with the highest SFRs ($\dot M_* \approx 50$--300~\Msolyr, with a median SFR of 94~\Msolyr), agreeing well with the inferred SFRs of observed \smm sources at $z \lesssim 3$. However, the model \smmbright galaxies in the tail-end of the redshift distribution ($z > 3$) had SFRs that were a factor of $\approx 3$ lower than those inferred by the observations, yet, we note that less than a fifth of the \smmbright model galaxies lie at these redshifts. Additionally, the redshift distribution of the galaxies in the \smmbright sample reasonably reproduced the shape and median value of the observed distribution. Finally, many of the integrated properties for the \smmbright model galaxies broadly agreed with the observed values of \smm sources. In combination, these results give us the confidence to use the \smmbright sample in an attempt to answer the questions relating to the origin, evolution and eventual fate of the observed \smm   population.

The \smmbright galaxies were found to be typically massive ($M_* \sim 10^{11}$~\Msol), star-forming ($\dot M_* \approx 100$~\Msolyr) galaxies at $z \approx 2$--3 with considerable dust masses ($M_{\mathrm{dust}} \sim 10^{8}$~\Msol) and gas fractions ($f_{\mathrm{gas}} \approx 50$\%) that host black holes that are undermassive for galaxies of their stellar masses. The $z=0$ descendants of the \smmbright galaxies were again massive ($M_* \geq 10^{11}$~\Msol), with all of the galaxies at $z=0$ with stellar masses above $M_* \geq 3 \times 10^{11}$~\Msol having had a \smmbright phase in their past. However, at a given stellar mass, the \smmbright galaxies were not obviously distinguishable from the general population of model galaxies. As the black holes of the once \smmbright galaxies do not remain undermassive by $z=0$, and come to lie on or around the median value for their stellar mass by the present day, they must of experienced a vigorous period of black hole growth (delayed after the starbursting phase) which grew the black hole and lowered the gas and dust content via AGN feedback. The typical life cycle of model SMGs, and their connection to black hole growth, will be the subject of future work.

The galaxies from the \smmbright sample had major merger (\mergerratio $\geq \frac{1}{4}$) fractions similar to those of the general population, suggesting that major mergers are not the primary driver of the model \smm population. This result would conflict with the findings of many observational studies that have suggested $\approx$ 100\% major merger fractions for \smm galaxies \citep[e.g.,][]{Ivison2007,Tacconi2008,Engel2010,Alaghband_Zadeh2012,Chen2015}. However, the concept of galaxies \squotes{in the state of a merger} is poorly defined; observations must deduce this state from a galaxy's asymmetry/disturbance or relative distance to a close companion, whereas we define it as being within a fixed number of dynamical times from coalescence. Furthermore, major mergers may play a greater role in triggering the most extreme (\fsmm $\gtrsim 5$~mJy) SMGs, which are those primarily studied observationally (however our current simulation volume is too small to fully explore this regime). Regardless of the definition, a mass/redshift matched control sample is essential to begin to infer the \squotes{importance} of mergers as triggering mechanisms, yet, given the natural abundance of mergers within a hierarchical formation of structure, even then disentangling their true importance can remain challenging. As the merger fractions of the \smmbright galaxies were so similar to the merger fractions of the general population, we would argue that mergers and interactions may be sufficient, but not necessary, to create a \smm starburst. Instead, the factor which appears essential is a significant gas reservoir due to an undermassive black hole.

The duration of the \squotes{\smm phase} for the \smmbright galaxies was found to be relatively brief (100--400~Myr) and contributed an average of $\approx 10$--20\% towards the total in situ stellar mass production of the model galaxies. The importance of the \smm phase for building up the stellar mass within a galaxy is an interesting question that is often raised when discussing the nature of \smm sources. Given the levels of star formation that are inferred for the observed \smm population (100s to 1000s of \Msolyr), the stellar mass of a massive galaxy ($M_* \gtrsim 10^{11}$~\Msol) could be formed in its totality in just a few hundred million years. This has spawned the suggestion that \smm galaxies are the progenitors of today's most massive spheroidal galaxies \citep[e.g.,][]{Lilly1999,Swinbank2006,Fu2013,Simpson2014}. Indeed, \citet{Swinbank2006} demonstrated that by adopting a \smm-burst lifetime of 300~Myr, coupled with the observed properties of SMGs at $z \approx 2$, the stellar populations of \smm galaxies could evolve onto the scaling relations of the most massive elliptical galaxies in the local Universe \citep[see also][]{Simpson2014}. The alternate extreme has also been suggested, whereby the \smm phase makes no significant contribution ($\approx 2$\%) to the eventual stellar mass of the SMG descendants at $z=0$ \citep{Gonzalez2011}. However, our results suggest the correct answer lies somewhere between these two extremes.

In contrast to the \smmbright galaxies, the highly star-forming \smmfaint galaxies were found to be a set of lower-mass ($M_* \sim 10^{10}$~\Msol) higher-redshift ($z>4$) galaxies (again with high dust masses, high gas fractions and undermassive black holes) that show stronger evidence of being driven primarily by major mergers. A similar fraction of the in situ stellar mass budget is built within the starbursting phase compared with the \smmbright galaxies ($\approx 10$--20\%). Their descendants at $z=0$, whilst also massive ($M_* \gtrsim 10^{11}$~\Msol), were typically metal-poor, alpha-enhanced and hosted overmassive black holes for galaxies of their stellar masses, potentially making them identifiable in today's parameter space.

To summarise, whilst collectively the galaxies in the two samples have the highest SFRs in the simulation and are the progenitors of the most massive galaxies at $z \approx 0$ ($M_* > 10^{11}$~\Msol), they each preferentially select galaxies in two different regimes. The \smmbright galaxies are massive, dust and gas rich star-forming galaxies at $z \approx 2$--3 with undermassive black holes, whereas the highly star-forming \smmfaint galaxies are high redshift, lower-mass, gas and dust rich starbursting galaxies which are more frequently triggered via a major merger. These high-redshift galaxies are likely missed by the current \smm surveys due to their higher dust temperatures. 

We have used the \eagle cosmological hydrodynamical simulation of galaxy formation to investigate the nature of the model galaxies with the highest mock \smm fluxes (\fsmm $\geq 1$~mJy). In addition, we investigated the nature of the \squotes{highly star-forming} model galaxies which were also \squotes{\smmfaint} (i.e., $\dot M_* \geq 80$~\Msolyr but \fsmm $< 1$~mJy). Here we report our main conclusions: 

\subsection{\smmbright galaxies}
\label{sect:conc_bright}
\begin{itemize}
    \item \textbf{The \smmbright model galaxies broadly reproduce the properties of the observed \smm population}. They have high SFRs ($\dot M_* \approx 50$--300~\Msolyr, see \cref{fig:sfr_distribution}), broadly reproduce the shape and median value of the observed redshift distribution (see \cref{fig:redshift_distribution}), and reproduce a variety of integrated galaxy and halo properties from the current observations of the \smm population (see \cref{fig:smg_properties_at_time}). 
    
    \item \textbf{The integrated properties of the \smmbright galaxies evolve with redshift.} At a given redshift, the \smmbright model population comprises of massive ($M_* \sim 10^{11}$~\Msol), gas ($f_{\mathrm{gas}} \approx 50$\%) and dust-rich ($M_{\mathrm{dust}} \sim 10^{8}$~\Msol) starbursting galaxies ($\dot M_* \approx 100$~\Msolyr) that host undermassive black holes (see \cref{fig:general_population}). With decreasing redshift, the \smmbright population have higher halo, stellar and black hole masses, they become increasingly gas poor, their sizes, velocity dispersions and dust masses increase, their metallicities remain approximately constant, and their dust temperatures decrease (see \cref{fig:smg_properties_at_time}). 
    
    \item \textbf{\smmbright galaxies have major merger (\mergerratio $\geq \frac{1}{4}$) fractions similar to the general population.} This would suggest that major mergers are not the primary trigger of the \eagle \smm population (see \cref{fig:merger_fraction}). Instead, what is critical is that there is an adequate gas reservoir present.
    
    \item \textbf{The majority of star production in \smmbright galaxies occurs outside the high SFR event(s)}. On average, 10--20\% of the stars that are born within the \smmbright model galaxies do so at high SFRs ($\dot M_* \geq 80$~\Msolyr), which implies a typical starbursting duration of $\approx 100$--400~Myr (combined between one or more events). Therefore, the majority of the stellar mass build-up in \smmbright galaxies occurs at lower SFRs, over longer periods of time (see \cref{fig:mass_in_burst}).
    
    \item \textbf{The descendants of the \smmbright population at $z=0$ are massive ($M_* > 10^{11}$~\Msol).} However, not all massive galaxies today were once \smmbright (see \cref{fig:cumulative_fraction}). At a given stellar mass, the descendants of \smmbright galaxies have gas fractions, dust masses, black hole masses, metallicities and alpha-enhancements (and all other integrated properties that we have explored) that are similar to the median trend for all galaxies (see \cref{fig:descendants}). This suggests that it would be difficult to identify the descendants \smmbright galaxies in today's parameter space. A moderate fraction of the once \smmbright galaxies evolve to become satellites by the present day (35\%, up from 10\% at the time they were selected) suggesting the environment may play a role in their formation.  
    
\end{itemize}

\subsection{Highly star-forming \smmfaint galaxies (i.e., $\dot M_* \geq 80$~\Msolyr but \fsmm $< 1$~mJy)}
\label{sect:conc_faint}
\begin{itemize}

    \item \textbf{Highly star-forming \smmfaint galaxies predominantly exist at higher redshift ($z >4$, see \cref{fig:redshift_distribution}).} Similar to the \smmbright population, highly star-forming \smmfaint galaxies have high SFRs, dusts masses and gas fractions, and host undermassive black holes (see \cref{fig:general_population}). The integrated properties of the highly star-forming \smmfaint galaxies evolve with redshift in a similar manner to the \smmbright galaxy population (see \cref{fig:smg_properties_at_time}).   

    \item \textbf{Highly star-forming \smmfaint galaxies galaxies have major merger (\mergerratio $\geq \frac{1}{4}$) fractions typically greater than the general population.} This would suggest that major mergers are more important for triggering this galaxy subset, particularly at lower stellar masses ($M_* \ll 10^{11}$~\Msol), where the enhancement in the major merger fraction is a factor of $\approx 3$ greater than those of the general population (see \cref{fig:merger_fraction}). 

    \item \textbf{The descendants of highly star-forming \smmfaint galaxies at $z=0$ have overmassive black holes, are metal-poor, and are alpha-enhanced for their stellar masses, relative to the median trends of the general population (see \cref{fig:descendants})}. In addition, many of the once highly star-forming \smmfaint galaxies evolve to become satellites of the most massive haloes within the simulation by the present day (32\% of the descendants evolved to become satellites, up from 3\% at the time they were highly star-forming) suggesting the environment may play a role in their formation.

    \item \textbf{Highly star-forming \smmfaint galaxies are faint in the \smm wavebands due to their higher dust temperatures.} This means that a large fraction of the highly star-forming galaxies within the Universe could be potentially missed by the current \smm surveys (see \cref{fig:smg_properties_at_time,fig:SED}), and also suggests that current observational works may be underestimating the total contribution to the cosmic SFR density above $z \gtrsim 3$ (as they would be missed in both the UV and \smm bands).

\end{itemize}

\section*{Acknowledgements}

SM thanks Michelle Furlong and Jonathan Davies for their many contributions to this work.

This work was supported by the Science and Technology Facilities Council (grant number ST/P000541/1) and the Academy of Finland (grant number 314238). RAC is a Royal Society University Research Fellow. IRS acknowledges support from the ERC Advanced Grant DUSTYGAL (321334).

This work used the DiRAC Data Centric system at Durham University, operated by the Institute for Computational Cosmology on behalf of the STFC DiRAC HPC Facility (\url{www.dirac.ac.uk}). This equipment was funded by BIS National E-infrastructure capital grant ST/K00042X/1, STFC capital grant ST/H008519/1, and STFC DiRAC Operations grant ST/K003267/1 and Durham University. DiRAC is part of the National E-Infrastructure.
\bibliographystyle{mnras}

\bibliography{mybib}
\begin{table*}

\caption{The median values and the 1$\sigma$ uncertainties (obtained via bootstrap resampling) for the integrated galaxy properties of the \eagle \smmbright model galaxies shown in \cref{fig:smg_properties_at_time}. From left to right: The halo mass, stellar mass, central supermassive black hole mass, total gas fraction $M_{\mathrm{gas}} / M_{\mathrm{gas+stars}}$, stellar half mass radius, stellar velocity dispersion, dust temperature, dust mass and stellar metallicity. The quoted errors for the redshifts represent the bin width.}

\begin{tabular}{crrrrrrrrr} \hline

$z$ & $M_{\mathrm{halo}}$ & $M_{\mathrm{*}}$ & $M_{\mathrm{BH}}$ & $f_{\mathrm{gas}}$ & HMR$_*$ & $\sigma_*$ & $T_{\mathrm{dust}}$ & $M_{\mathrm{dust}}$ & $Z_*$ \\

& [$10^{12}$M$_{\odot}$] & [$10^{10}$M$_{\odot}$] & [$10^{7}$M$_{\odot}$] & [\%] & [pkpc] & [km~s$^{-1}$] & [K] & [$10^{8}$M$_{\odot}$] & [$10^{-3}$] \\

\hline\hline

$1.3\pm0.3$&$17.6^{+4.7}_{-9.6}$&$13.2^{+3.8}_{-2.8}$&$19.0^{+11.7}_{-7.0}$&$29.5^{+4.3}_{-4.3}$&$5.6^{+1.4}_{-7.6}$&$251^{+16}_{-13}$&$29.8^{+1.5}_{-0.1}$&$2.5^{+0.1}_{-0.4}$&$17.4^{+3.3}_{-0.9}$\\
$1.8\pm0.2$&$12.6^{+3.0}_{-2.8}$&$10.1^{+1.7}_{-3.1}$&$20.7^{+18.3}_{-1.8}$&$36.6^{+3.0}_{-1.4}$&$4.0^{+0.2}_{-3.2}$&$247^{+24}_{-8}$&$28.5^{+0.7}_{-1.0}$&$2.8^{+0.2}_{-0.3}$&$15.7^{+2.2}_{-3.7}$\\
$2.2\pm0.1$&$9.1^{+4.8}_{-0.2}$&$8.7^{+0.1}_{-2.6}$&$15.4^{+9.5}_{-1.4}$&$37.4^{+3.2}_{-1.7}$&$2.5^{+0.4}_{-0.5}$&$223^{+25}_{-17}$&$28.5^{+0.4}_{-0.9}$&$2.5^{+0.2}_{-0.5}$&$17.9^{+0.1}_{-1.8}$\\
$2.4\pm0.1$&$6.8^{+2.1}_{-0.7}$&$8.6^{+0.5}_{-2.2}$&$8.0^{+2.1}_{-5.5}$&$40.6^{+2.9}_{-2.6}$&$1.3^{+0.1}_{-0.5}$&$209^{+5}_{-16}$&$31.7^{+1.6}_{-1.3}$&$2.3^{+0.2}_{-0.1}$&$18.8^{+1.2}_{-0.9}$\\
$2.8\pm0.3$&$3.9^{+0.6}_{-2.7}$&$6.7^{+0.5}_{-0.4}$&$4.8^{+1.5}_{-1.1}$&$44.6^{+0.3}_{-0.6}$&$1.6^{+0.2}_{-0.7}$&$188^{+5}_{-14}$&$31.2^{+1.1}_{-1.1}$&$2.2^{+0.3}_{-0.1}$&$16.0^{+1.3}_{-3.0}$\\
$3.8\pm0.7$&$2.3^{+0.1}_{-0.3}$&$6.1^{+0.7}_{-0.1}$&$3.8^{+0.2}_{-7.9}$&$51.2^{+1.2}_{-3.7}$&$0.7^{+0.1}_{-0.1}$&$199^{+8}_{-4}$&$32.6^{+1.5}_{-0.8}$&$1.7^{+0.0}_{-0.3}$&$18.3^{+0.8}_{-0.4}$\\

\hline
\end{tabular}
\label{tab:prop_smm_bright}
\end{table*}

\begin{table*}
\caption{The same as \cref{tab:prop_smm_bright}, but now for the \eagle highly star-forming \smmfaint galaxies.}

\begin{tabular}{crrrrrrrrr} \hline

$z$ & $M_{\mathrm{halo}}$ & $M_{\mathrm{*}}$ & $M_{\mathrm{BH}}$ & $f_{\mathrm{gas}}$ & HMR$_*$ & $\sigma_*$ & $T_{\mathrm{dust}}$ & $M_{\mathrm{dust}}$ & $Z_*$ \\

& [$10^{12}$M$_{\odot}$] & [$10^{10}$M$_{\odot}$] & [$10^{7}$M$_{\odot}$] & [\%] & [pkpc] & [km~s$^{-1}$] & [K] & [$10^{8}$M$_{\odot}$] & [$10^{-3}$] \\

\hline\hline

$1.7\pm0.6$&$11.6^{+3.3}_{-2.4}$&$7.5^{+1.4}_{-3.5}$&$11.8^{+3.1}_{-9.4}$&$43.1^{+7.4}_{-5.4}$&$10.9^{+4.0}_{-2.3}$&$228^{+11}_{-17}$&$29.6^{+1.7}_{-1.0}$&$1.9^{+0.1}_{-0.1}$&$12.3^{+0.6}_{-0.4}$\\
$2.7\pm0.4$&$2.5^{+0.5}_{-0.0}$&$3.5^{+0.3}_{-0.4}$&$3.6^{+0.6}_{-1.1}$&$62.6^{+1.1}_{-4.3}$&$1.3^{+0.5}_{-1.2}$&$163^{+1}_{-123}$&$33.0^{+1.5}_{-2.9}$&$1.3^{+0.1}_{-0.3}$&$10.0^{+0.4}_{-0.8}$\\
$3.4\pm0.3$&$2.1^{+0.5}_{-0.0}$&$2.5^{+0.4}_{-1.2}$&$1.2^{+0.0}_{-2.5}$&$68.1^{+3.3}_{-5.0}$&$1.2^{+0.3}_{-0.8}$&$176^{+25}_{-0}$&$33.3^{+0.1}_{-1.9}$&$1.0^{+0.0}_{-0.2}$&$9.9^{+1.8}_{-2.1}$\\
$4.1\pm0.4$&$1.4^{+0.2}_{-0.1}$&$2.3^{+0.3}_{-0.7}$&$1.1^{+0.5}_{-0.5}$&$70.0^{+6.0}_{-5.7}$&$0.6^{+0.1}_{-0.2}$&$164^{+2}_{-7}$&$37.6^{+0.7}_{-0.2}$&$0.7^{+0.0}_{-0.0}$&$13.6^{+3.2}_{-1.3}$\\
$4.6\pm0.0$&$1.4^{+0.5}_{-0.2}$&$2.9^{+0.7}_{-0.5}$&$4.2^{+3.3}_{-5.4}$&$66.0^{+0.3}_{-2.5}$&$0.5^{+0.0}_{-0.1}$&$169^{+7}_{-7}$&$38.0^{+1.9}_{-1.7}$&$0.8^{+0.1}_{-0.1}$&$14.2^{+0.8}_{-0.6}$\\
$4.8\pm0.3$&$1.1^{+0.0}_{-0.0}$&$1.9^{+0.1}_{-0.3}$&$0.4^{+0.1}_{-0.1}$&$76.0^{+5.0}_{-0.3}$&$0.3^{+0.0}_{-0.1}$&$145^{+5}_{-9}$&$40.8^{+0.1}_{-1.7}$&$0.4^{+0.0}_{-0.0}$&$10.1^{+0.4}_{-4.9}$\\
$5.1\pm0.0$&$1.1^{+0.2}_{-0.0}$&$2.0^{+0.2}_{-0.2}$&$0.6^{+0.2}_{-0.7}$&$69.2^{+2.2}_{-4.5}$&$0.4^{+0.0}_{-0.1}$&$150^{+3}_{-8}$&$41.3^{+1.5}_{-0.2}$&$0.4^{+0.0}_{-0.0}$&$12.5^{+2.5}_{-1.9}$\\
$5.6\pm0.4$&$0.9^{+0.1}_{-0.0}$&$1.6^{+0.2}_{-0.1}$&$0.1^{+0.1}_{-1.3}$&$76.4^{+4.3}_{-1.3}$&$0.4^{+0.1}_{-0.1}$&$168^{+16}_{-5}$&$39.9^{+1.2}_{-1.4}$&$0.3^{+0.1}_{-0.1}$&$11.7^{+2.2}_{-0.6}$\\

\hline
\end{tabular}
\label{tab:prop_smm_faint}
\end{table*}

\bsp	
\label{lastpage}

\end{document}